\documentclass[aps, prd, twocolumn, notitlepage, nofootinbib]{revtex4-2}
\usepackage{graphicx}
\usepackage{float}
\usepackage{amsmath}
\usepackage{color}
\usepackage{tensor}
\usepackage{epstopdf}
\usepackage{xcolor}
\usepackage{wasysym}
\usepackage{hyperref}
\hypersetup{
    colorlinks,
    linkcolor={blue!50!black},
    citecolor={red!50!black},
    urlcolor={blue!80!black}
}

\usepackage{multibib}
\newcommand{\be}{\begin{equation}}
\newcommand{\ee}{\end{equation}}
\newcommand{\ba}{\begin{eqnarray}}
\newcommand{\ea}{\end{eqnarray}}

\newcommand{\beq}{\begin{equation}}
\newcommand{\eeq}{\end{equation}}
\newcommand{\beqa}{\begin{eqnarray}}
\newcommand{\eeqa}{\end{eqnarray}}
\newcommand{\nn}{\nonumber}
\graphicspath{{plots/}}

\begin{document}

\title{Timelike orbits around accelerating black holes}

\author{Mohammad Bagher Jahani Poshteh}
\email{jahani@ipm.ir}
\affiliation{School of Physics, Institute for Research in Fundamental Sciences (IPM), P.O. Box 19395-5531, Tehran, Iran}

%
%\date{\today}

%\pacs{04.50.Gh, 04.70.-s, 05.70.Ce}

\begin{abstract}
	We study the geodesics of massive particles around accelerating Schwarzschild black hole.
	We show that the radius of the innermost stable circular orbit increases and the angular momentum of particle at this orbit decreases by increasing the acceleration. Apart from quantitative influence, the acceleration qualitatively changes the physics. We show that in accelerating black hole spacetime there exist an outermost stable circular orbit in flat, de Sitter, and anti de Sitter background.
	Investigations of radial geodesics show that the acceleration acts like a repulsive force in the sense that test particles around accelerating black holes can move radially outward, unless there exist a large negative value of cosmological constant in the background to compensate the repulsive force.
	We also investigate the precession of perihelion of orbits around accelerating black holes. The precession would be larger compared to the non-accelerating case. It is also shown that the precession in anti de Sitter background could be opposite to the particle motion.
\end{abstract}

\maketitle

\section{Introduction}

The study of relativistic motion of particles in gravitational field began short after the general theory of relativity was introduced~\cite{einstein1916,Dyson,Einstein}. In light of the investigations of lightlike geodesics we have found many interesting phenomena, e.g.~gravitational lensing \cite{Schneider,Darwin59,Darwin61,poshteh2019,Ellis,Virbhadra} and black hole shadow \cite{synge,hennigar2018,akiyama2019}. These features are in common among black holes and naked singularities \cite{virbhadra2002,joshi2020} and ultra-compact objects \cite{cunha2018shadow,shaikh2019}.

The investigation of timelike geodesics around massive objects could help us understand the nature of the objects. In this paper we are interested in timelike geodesics in accelerating black hole spacetime. The motion of massive test particles around non-accelerating black holes in general relativity has been thoroughly studied in \cite{chandrasekhar1998}. In \cite{wilkins1972,glampedakis2002,fujita2009} fundamental frequencies of particles' motion around Kerr black holes have been investigated. Circular motions on equatorial plane around Kerr–Newman black holes and Kerr–Newman naked singularities are compared in \cite{pugliese2013}. In \cite{Shoom:2015slu} null and timelike geodesics on equatorial plane of a distorted Schwarzschild black hole are studied. In \cite{Bambhaniya:2019pbr}, for a class of naked singularity spacetimes, it has been shown that the perihelion precession can be in the opposite direction of particle's motion. Precession of timelike bound orbits in Kerr spacetime has also been studied and it is found that the precession is positive \cite{Bambhaniya:2020zno}.

Black holes can be pair produced on cosmic strings \cite{hr,eardley1995,ashoorioon2014} as well as in a de Sitter \cite{mellor1989,mann1995,dias2004:2} or magnetic field \cite{garfinkle1991,hawking1995,dowker1994} background (see \cite{ashoorioon2021} for the black hole production rate on cosmic string in a de Sitter space with background magnetic field). On the other hand, we might have primordial black holes got attached to cosmic strings in the early Universe~\cite{vilenkin1985}. All of these black holes will be accelerating (due to the tension of the cosmic string~\cite{vilenkin1985} and/or the force exerted by the positive cosmological constant and/or the magnetic field \cite{ashoorioon2021}).

Accelerating black holes could evolve to supermassive black holes~\cite{vilenkin2018} (see also~\cite{gussmann2021}). However, if these black holes have played a role in structure formation, their velocity should be small~\cite{vilenkin2018}; which means that the acceleration should be small. It has been speculated, from observational point of view, that the black hole at the center of our Galaxy is connected to cosmic string~\cite{morris2017}.

Study of null geodesics around accelerating black hole has recently attracted some attention \cite{lim2021,frost2021,Ashoorioon:2021znw}. The shadow of the accelerating black hole has been studied in \cite{grenzebach2015,zhang2021}. On the other hand, thermodynamics of these black holes have been investigated in \cite{appels2016,anabalon2018,anabalon2019}. Near horizon symmetries of accelerating black holes have also been studied \cite{brenner2021}.

Accelerating black holes are described by C metric. In \cite{pravda2001}, for the particles co-accelerating with the black hole at constant distance, null and timelike geodesics are found in the standard $\{x, y\}$ coordinates \cite{kinnersley1970} as well as Weyl coordinates \cite{bonnor1983} and the coordinates adapted to the boost-rotation symmetry \cite{k1989}. Generalization to AdS C metric \cite{chamblin2001} and spinning C metric \cite{pravda2002} have also been done. In \cite{lim2014}, by studying the effective potential of test particles around the black hole, accessible regions for null/timelike geodesics are presented. Stability of geodesics is also discussed.

The aim of this paper is to study the motion of massive particles in accelerating black hole spacetime with cosmological constant. We would like to study radial geodesics as well as the innermost stable circular orbit (ISCO) of particles around accelerating supermassive black holes. We would also like to investigate the perihelion precession of such orbits. In our examples we consider a non-rotating black hole with the same mass as the black hole at the center of Milky Way Galaxy, Sgr A*. We also use the orbital data of S2 star around Sgr A*. Considering the upper bound on the acceleration of supermassive black holes~\cite{vilenkin2018}, we show that we can take S2 to be nearly on equatorial plane of accelerating black hole
during one complete period.

Study of radial motion of test particles in accelerating black hole spacetime shows that these particles can move radially outward. This means that accelerating black hole exert some sort of repulsive force on test particles. We also find the interesting result that there exist an upper bound on the radius of stable circular orbits around the accelerating black hole. These features are in common among accelerating black holes and black holes in de Sitter spacetime.

We also investigate the precession of orbits around accelerating black holes. We show that the precession of perihelion is larger for larger values of black hole acceleration. However, for black holes in anti de Sitter background, the precession can be negative (the orbit precesses in the opposite direction of the motion).

The outline of our paper is as follows. In the next section we present the Lagrangian of motion on a plane perpendicular to the direction of acceleration (equatorial plane) and study the radial motion on this plane. Stable circular orbits on the equatorial plane is studied in Sec.~\ref{sec:isco}. In Sec.~\ref{sec:precession} we study the precession of the orbits. We conclude our paper in Sec.~\ref{sec:con}. We work in geometric units where $G=c=1$ and use mostly-positive signature for the spacetime metric.

\section{Radial motions on equatorial plane}\label{sec:equtorial}

It has been shown in~\cite{vilenkin2018} that the velocity of supermassive black holes connected to cosmic string should be less than $100\, {\rm km}/{\rm s}$ so that they could be captured by galaxies during the structure formation. This, in turn, constrain tension of the cosmic string to $\mu \lesssim 10^{-19}\sqrt{M/M_{\astrosun}}$, where $M$ is the mass of supermassive black hole connected to the cosmic string. For the black hole at the center of our Galaxy, with $M_{{\rm Sgr A}^{*}}\simeq 4.23 \times 10^6 M_{\astrosun}$~\cite{johannsen2016}, the tension of the cosmic string attached to it would be $\mu \lesssim 2.06 \times 10^{-16}$. Therefore we find from the Newton second law, $\mu = \alpha M_{{\rm Sgr A}^{*}}$, that the acceleration is $\alpha \lesssim 3.29 \times 10^{-26} {\rm m}^{-1}$.

The orbital period of S2 star around Sgr A* is about 16 years~\cite{mnd,hees2017}. We take the acceleration of Sgr A* to be $\alpha = 10^{-26} {\rm m}^{-1}$ --- this is about the largest value allowed by~\cite{vilenkin2018}. Assuming the initial velocity of the black hole to be zero, the displacement of the black hole during the period of S2 is $2.29 \times 10^8 {\rm m}$. This is much smaller than the Schwarzschild radius of the black hole which is of the order of $10^{10} {\rm m}$. Therefore we can assume that, if S2 starts its period near the equatorial plane of the black hole, it will remain near this plane during one period of the orbit around the slowly accelerating black hole.

Spacetime around uniformly accelerating black holes is described by C metric.
\footnote{This spacetime is of algebraic type D. It has been proved in~\cite{demianski1980} that vacuum type D spacetimes admit Killing and Killing-Yano tensors if (and only if) they are without acceleration.}
(A)dS C metric has been studied in, for example, \cite{mann1995,dias2003,Podolsky:2000pp,Podolsky:2002nk,chen2015new}. For a black hole of mass $m$ and uniform acceleration $\alpha$, in a background spacetime with cosmological constant $\Lambda$, the metric can be written in the following form~\cite{griffiths2006new}
\begin{eqnarray}
	ds^2&=&\frac{1}{(1+\alpha r\cos\theta)^2} \nn\\
	&\times&\left[-Q(r)dt^2+\frac{dr^2}{Q(r)}+\frac{r^2d\theta^2}{P(\theta)}+P(\theta)r^2\sin^2\theta d\phi^2\right],\nn\\ \label{eqn:metric:griffiths}
\end{eqnarray}
where
\begin{eqnarray}
	Q(r)&=&(1-\alpha^2 r^2)\left(1-\frac{2m}{r}\right) - \frac{\Lambda}{3}r^2, \nn\\
	P(\theta)&=&1+2\alpha m \cos\theta. \label{eqn:tran:griffiths}
\end{eqnarray}
The coordinate $t$ ranges over all of ${\rm I\!R}$ and $0 \leq \theta \leq \pi$. The periodicity of $\phi$ is $-C_0\pi \leq \phi \leq C_0\pi$. One can choose $C_0$ so as to eliminate the conical singularity at one of the poles. We take $C_0=(1+2\alpha m)^{-1}$ to eliminate the singularity at $\theta = 0$ \cite{lim2014}.
\footnote{Setting $C_0=(1-2\alpha m)^{-1}$, on the other hand, eliminates the singularity at $\theta = \pi$ \cite{lim2014}. For other choices of $C_0$ neither the singularity at $\theta = 0$ nor the one at $\theta = \pi$ would be eliminated and one has cosmic strings at both poles.}

We would like to study the geodesics on the equatorial plane of the accelerating black holes. We assume that the direction of acceleration is perpendicular to this plane. On equatorial plane the metric \eqref{eqn:metric:griffiths} reduces to
\be
ds^2=-Qdt^2+\frac{dr^2}{Q}+r^2d\theta^2+r^2d\phi^2, \label{eqn:metric}
\ee
with the metric function given by Eq.~\eqref{eqn:tran:griffiths}.

\begin{figure}[htp]
	\centering
	\includegraphics[width=0.45\textwidth]{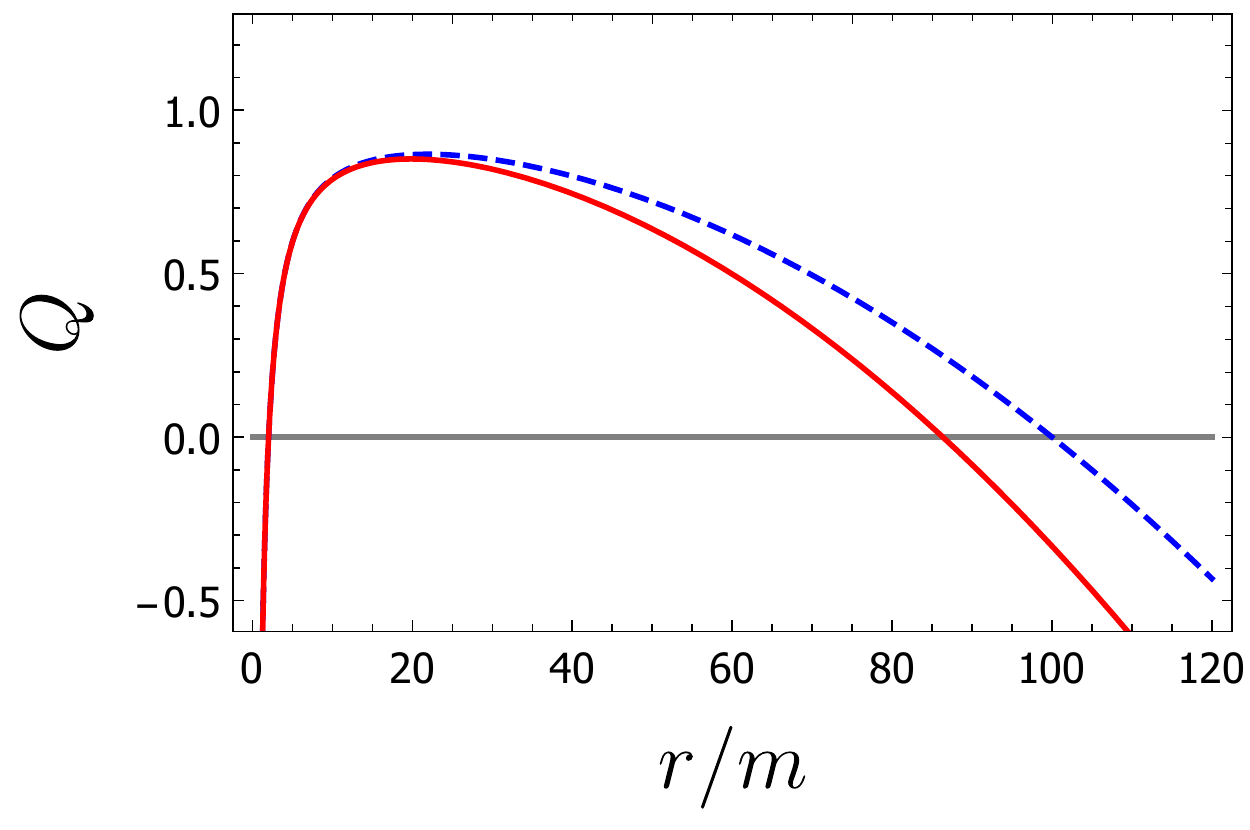}
	\includegraphics[width=0.45\textwidth]{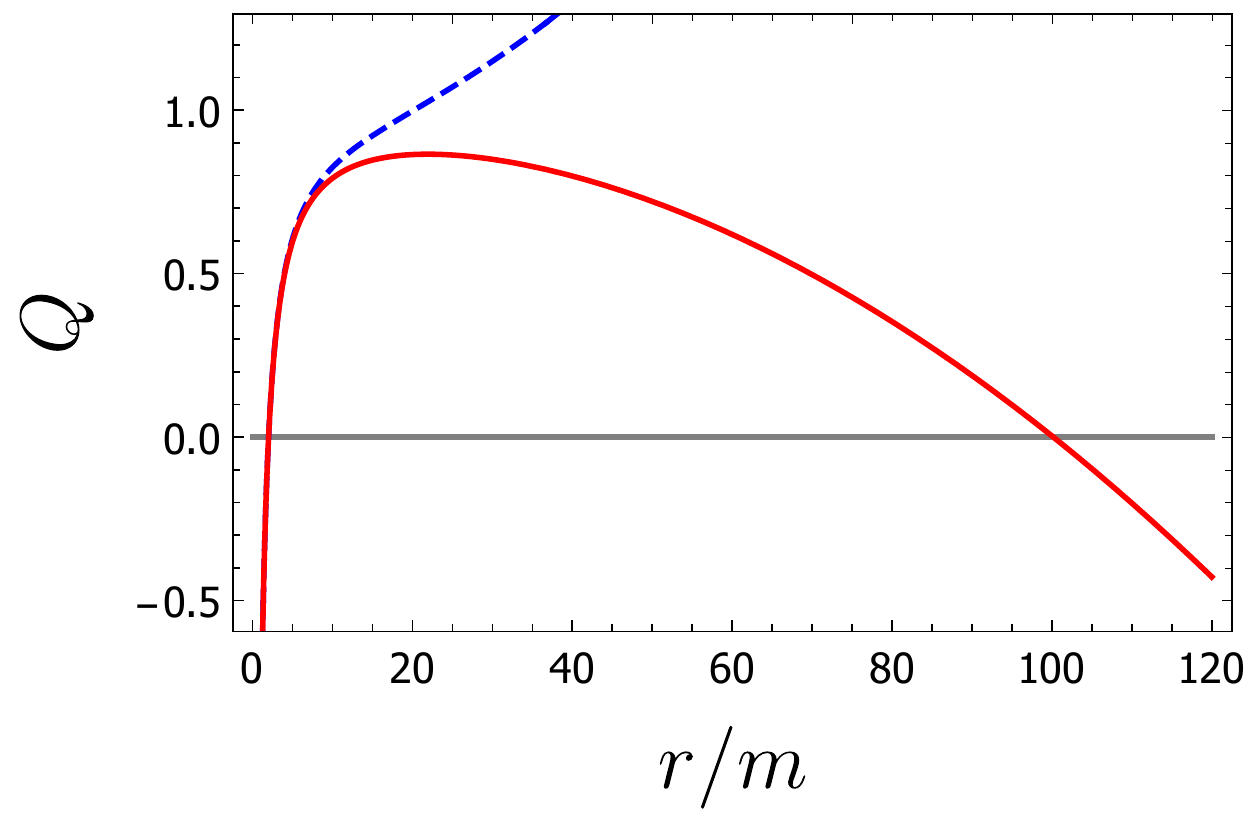}
	\caption{\textit{Top}: The metric function of an accelerating black hole in flat (dashed blue plot) and de Sitter with $m^2 \Lambda = 10^{-4}$ (solid red plot) background. \textit{Bottom}: The metric function of an accelerating black hole in anti de Sitter background with $m^2 \Lambda = -10^{-3}$ (dashed blue plot) and $m^2 \Lambda = -10^{-6}$ (solid red plot). We have taken $m \alpha = 10^{-2}$.}
	\label{fig:metric}
\end{figure}

The metric function $Q$ has been plotted in Fig.~\ref{fig:metric}. We have taken some large values of acceleration and cosmological constant so that we could better illustrate the behavior of the metric function in different cases. For the case of flat and de Sitter background the metric function has two zeros for positive $r$ and has a maximum between them. The behavior of the metric function in the case of anti de Sitter background is more interesting. For $\Lambda > -3 \alpha^2$ there are two zeros for positive $r$, however if $\Lambda < -3 \alpha^2$ there would be only one zero and the metric function goes to $+\infty$ as $r$ increases.
\footnote{For $\Lambda > -3 \alpha^2$ there is a maximum between two zeros of the metric function. As  $\Lambda \rightarrow (-3 \alpha^2)^{+}$ the point at which the maximum appears goes to $+\infty$.}

The region between the two zeros of $Q$ --- one associated to event horizon of the black hole, the other to acceleration horizon --- is the domain of outer communication in which $Q > 0$. For $\Lambda < -3 \alpha^2$ this domain contains any radius larger than the radius of black hole's event horizon. The motion of test body around the black hole is in the region of outer communication. Therefore we only consider the range of $r$ which is in the domain of outer communication.

The Lagrangian governing the geodesic motion is $2\mathcal{L}=g_{\gamma\sigma}\dot{x}^\gamma\dot{x}^\sigma=-\mu^2$, where $\mu$ is the rest mass of the infalling particle and dot, for timelike geodesics, represents differentiation with respect to the proper time (for a detailed study of timelike geodesics in Schwarzschild and Kerr background see~\cite{chandrasekhar1998}). Using the metric \eqref{eqn:metric} the Lagrangian on equatorial plane would be
\begin{equation}
	\label{eqn:lag}
	\mathcal{L}=\frac{1}{2}\left(-Q\dot{t}^{2}+\frac{\dot{r}^{2}}{Q}+r^{2}\dot{\phi}^{2}\right).
\end{equation}

The canonical momenta $p_t=-\frac{\partial\mathcal{L}}{\partial\dot{t}}$ and $p_\phi=\frac{\partial\mathcal{L}}{\partial\dot{\phi}}$ are conserved because of the symmetry of the metric, i.e.~$\dot{p}_t=-\frac{\partial\mathcal{L}}{\partial t}=0$ and $\dot{p}_\phi=\frac{\partial\mathcal{L}}{\partial \phi}=0$. They are the energy and angular momentum of the infalling particle and can be found to be
\be
E=p_t=Q\dot{t}, \qquad L_z=p_\phi = r^2\dot{\phi}.
\ee
We note here that if the acceleration has a component parallel to the equatorial plane which is of the same order or larger than the perpendicular component, then the third component of the angular momentum is no longer a constant of motion. In this case geodesic equations cannot be analytically integrated.
By using these quantities we can rewrite Eq.~\eqref{eqn:lag} as
\begin{equation}
	\label{timelikegeo}
	\dot{r}^2=\tilde{E}^2-Q\left[1+\frac{\tilde{L}_{z}^{2}}{r^2}\right],
\end{equation}
where $\tilde{E}$ and $\tilde{L}_z$ are respectively the energy and angular momentum per unit rest mass and the dot now represents differentiation with respect to proper time per unit rest mass. Note also that the second term on the right hand side of Eq.~\eqref{timelikegeo},
\be
\tilde{V}=Q\left[1+\frac{\tilde{L}_{z}^{2}}{r^2}\right],
\ee
is the effective potential.

Now we consider radial motion in accelerating black hole spacetime.
\footnote{For a black hole accelerating in $z$ direction, we consider radial motion in $x-y$ plane.}
For radial geodesics we have $\tilde{L}_{z} = 0$. Therefore Eq.~\eqref{timelikegeo} reduces to
\be
\dot{r}^2=\tilde{E}^2-Q.
\ee
If a test particle is falling from rest at $r_0$ we find $\tilde{E}^2 = Q(r_0)$.

For non-accelerating Schwarzschild black hole in anti de Sitter background it has been shown that the test particle always plunges into the black hole \cite{Cruz:2004ts}. This is due to the attractive force generated by the negative cosmological constant. However, for non-accelerating Schwarzschild black hole in de Sitter background, it is known that the test particle can move radially outward/inward if $r_0$ is larger/smaller than a specific value \cite{jaklitsch1989particle}.

The acceleration of the test particle which is at rest at $r_0$ is given by \cite{jaklitsch1989particle}
\be
\ddot{r}=\frac{Q'(r)\left[\dot{r}^2-Q(r_0)\right]}{2 Q(r)}.
\ee
For $\Lambda > -3 \alpha^2$ the metric function $Q$ has two zeros at positive $r$ (see Fig.~\ref{fig:metric}). Let us denote the smaller root by $r_+$ and the larger one by $r_{++}$. Between these two roots the metric function has a maximum at $r_M$. Suppose a test particle starts its motion at $r_0$. For $r_+<r_0<r_M$ the slope of the metric function is positive, therefore the initial acceleration $\ddot{r}=-Q'(r_0)/2$ is negative. This means that the test particle plunges into the black hole. For $r_M<r_0<r_{++}$, however, the acceleration is positive and the test particle moves radially outward. On the other hand, in anti de Sitter background with $\Lambda < -3 \alpha^2$ the test particle always plunges into the event horizon.

Therefore we find that acceleration acts like a positive cosmological constant. Similarity between a positive cosmological constant and cosmic strings has already been claimed in the context of black hole pair production in the author's earlier work \cite{ashoorioon2021}. We also see that a negative cosmological constant can compensate the acceleration ($\Lambda = -3 \alpha^2$ completely compensate the effect of $\alpha$).

\section{Stable circular orbits}\label{sec:isco}

\begin{figure}[htp]
	\centering
	\includegraphics[width=0.45\textwidth]{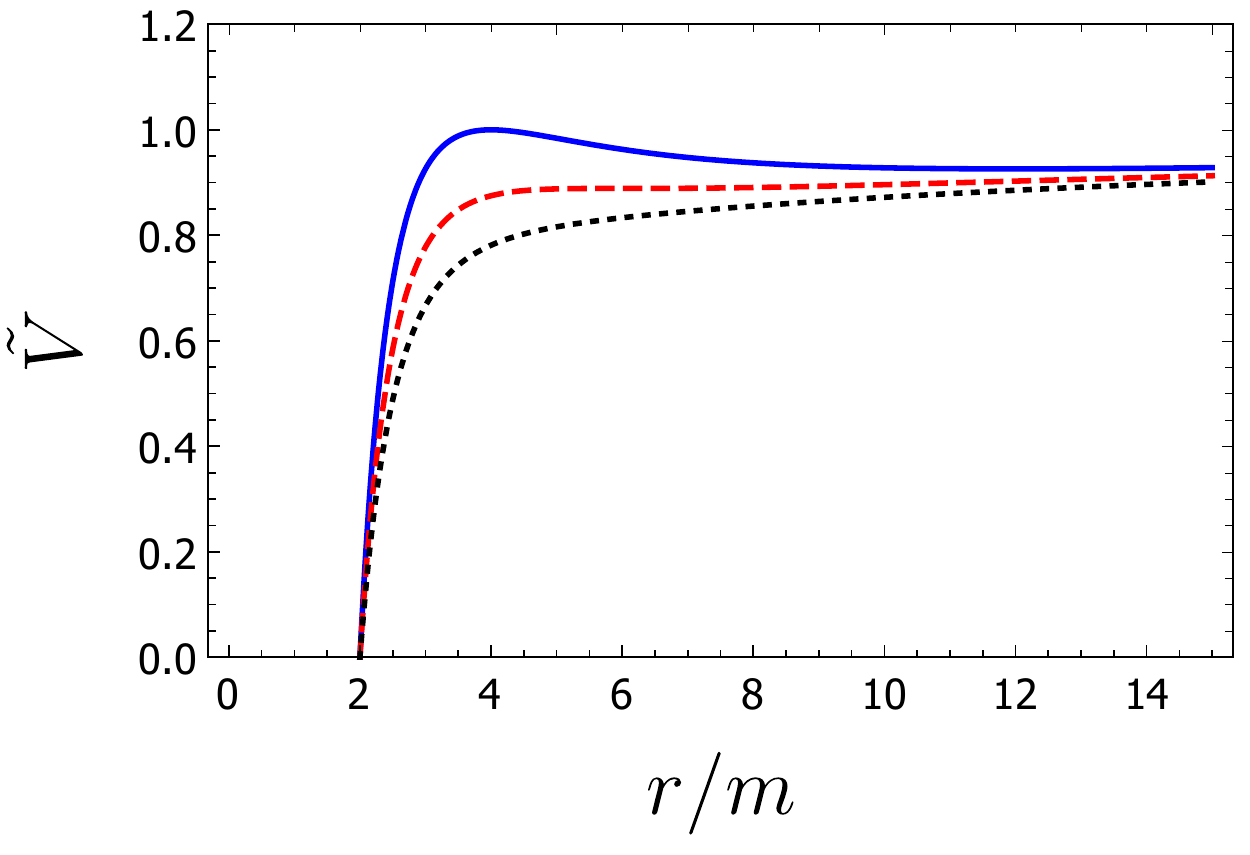}
	\caption{The effective potential for different values of angular momentum in the region near the black hole horizon. We have taken $m \alpha = 10^{-26}$, $m^2 \Lambda = 10^{-52}$, and $\tilde{L}_{z}=3 m$ (dotted black plot), $\tilde{L}_{z}=\tilde{L}_{z, {\rm ISCO}}\approx 3.4641 m$ (dashed red plot), and $\tilde{L}_{z}=4 m$ (solid blue plot). The maximum and minimum of the solid blue plot are respectively at $r \approx 4m$ and $r \approx 12m$. The inflection point of the dashed red plot is at $r \approx 6m$.}
	\label{fig:potential}
\end{figure}

Here we consider the circular geodesics for which $\tilde{L}_{z} \neq 0$. We have plotted the effective potential in Fig.~\ref{fig:potential} for different values of the angular momentum in the region near the event horizon of the black hole (we take approximately the observed value of the cosmological constant, $\Lambda \simeq 10^{-52} {\rm m}^{-2}$~\cite{Padmanabhan:2002ji}). In this region, we see that for $\tilde{L}_{z}$ slightly above $\tilde{L}_{z, {\rm ISCO}}$ there are a minimum and a maximum in the plot of the potential which are associated to stable and unstable circular orbits, respectively. The minimum gets closer to the black hole as $\tilde{L}_{z}$ decreases. For $\tilde{L}_{z} = \tilde{L}_{z, {\rm ISCO}}$ we have the ISCO which is at the inflection point of the effective potential.

\begin{figure}[htp]
	\centering
	\includegraphics[width=0.45\textwidth]{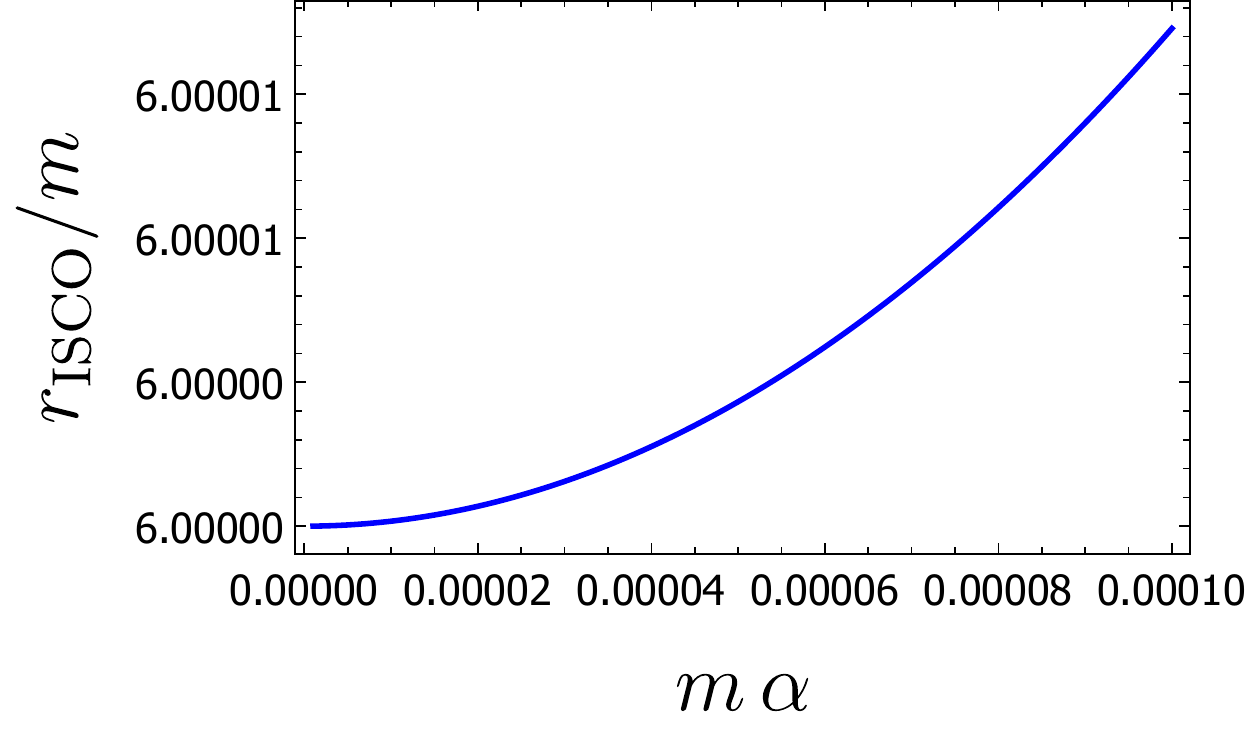}
	\includegraphics[width=0.45\textwidth]{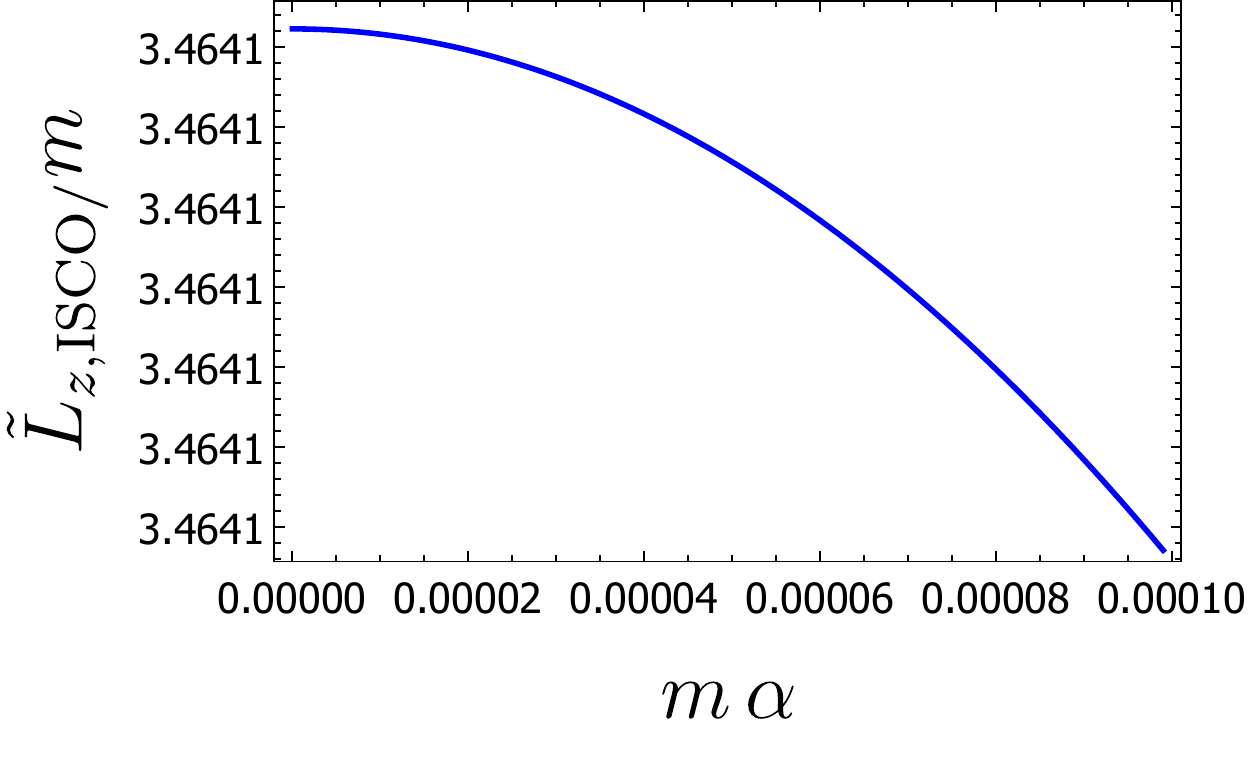}
	\caption{\textit{Top}: The radius of ISCO as a function of the acceleration of the black hole. \textit{Bottom}: The angular momentum of the particle on ISCO. We have taken $m^2 \Lambda = 10^{-52}$.}
	\label{fig:isco}
\end{figure}

We find that the radius of ISCO and the angular momentum of particle at this orbit depends on the acceleration. In Fig.~\ref{fig:isco}, using numerical techniques, we have plotted the radius of ISCO and the angular momentum at ISCO as a function of the acceleration. We see that the radius of ISCO increases and the angular momentum of particle at ISCO decreases by increasing the acceleration. This means that if the black hole is accelerating, the infalling particle can orbit on a stable circular orbit around the hole with a larger radius and smaller angular momentum.

\begin{figure}[htp]
	\centering
	\includegraphics[width=0.45\textwidth]{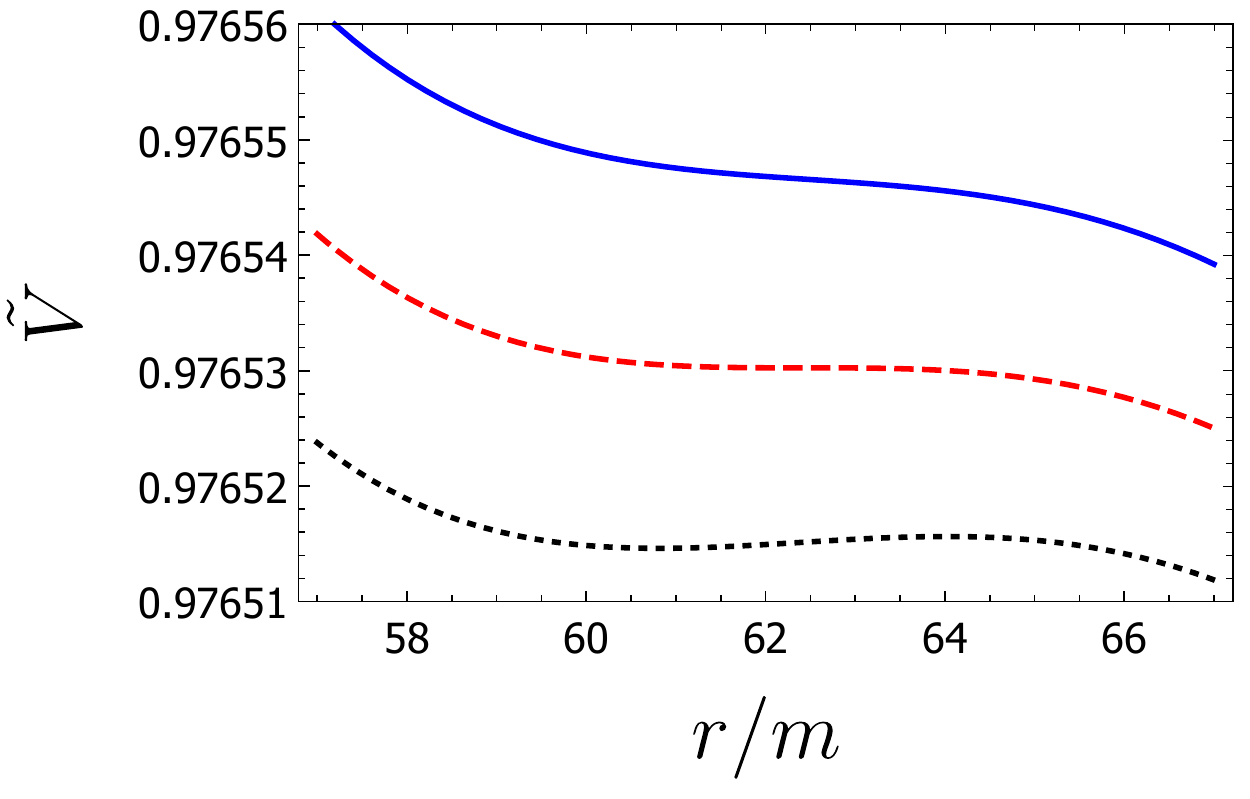}
	\caption{The effective potential for different values of angular momentum in the region far from the black hole horizon. We have taken $m \alpha = 10^{-3}$, $\Lambda = 0$, and $\tilde{L}_{z}=7.05800 m$ (dotted black plot), $\tilde{L}_{z}=\tilde{L}_{z, {\rm OSCO}}\approx 7.06232 m$ (dashed red plot), and $\tilde{L}_{z}= 7.06700 m$ (solid blue plot).}
	\label{fig:osco}
\end{figure}

For non-accelerating black holes, it is known that there exists an outermost stable circular orbit (OSCO) in a background spacetime with a positive cosmological constant \cite{howes1979existence,Boonserm:2019nqq}. We have find that, for accelerating black holes, OSCO exists for flat and AdS background as well as in dS spacetime. This point has been presented in Fig.~\ref{fig:osco}. Let us call the region in which the OSCO takes place as the far region. We see that for $\tilde{L}_{z}$ slightly below $\tilde{L}_{z, {\rm OSCO}}$ there are a minimum and a maximum in the plot of the potential which are, respectively, associated to stable and unstable circular orbits. The minimum goes away from the black hole as $\tilde{L}_{z}$ increases. For $\tilde{L}_{z} = \tilde{L}_{z, {\rm OSCO}}$ we have the OSCO which is at the inflection point of the effective potential.

\begin{figure}[htp]
	\centering
	\includegraphics[width=0.45\textwidth]{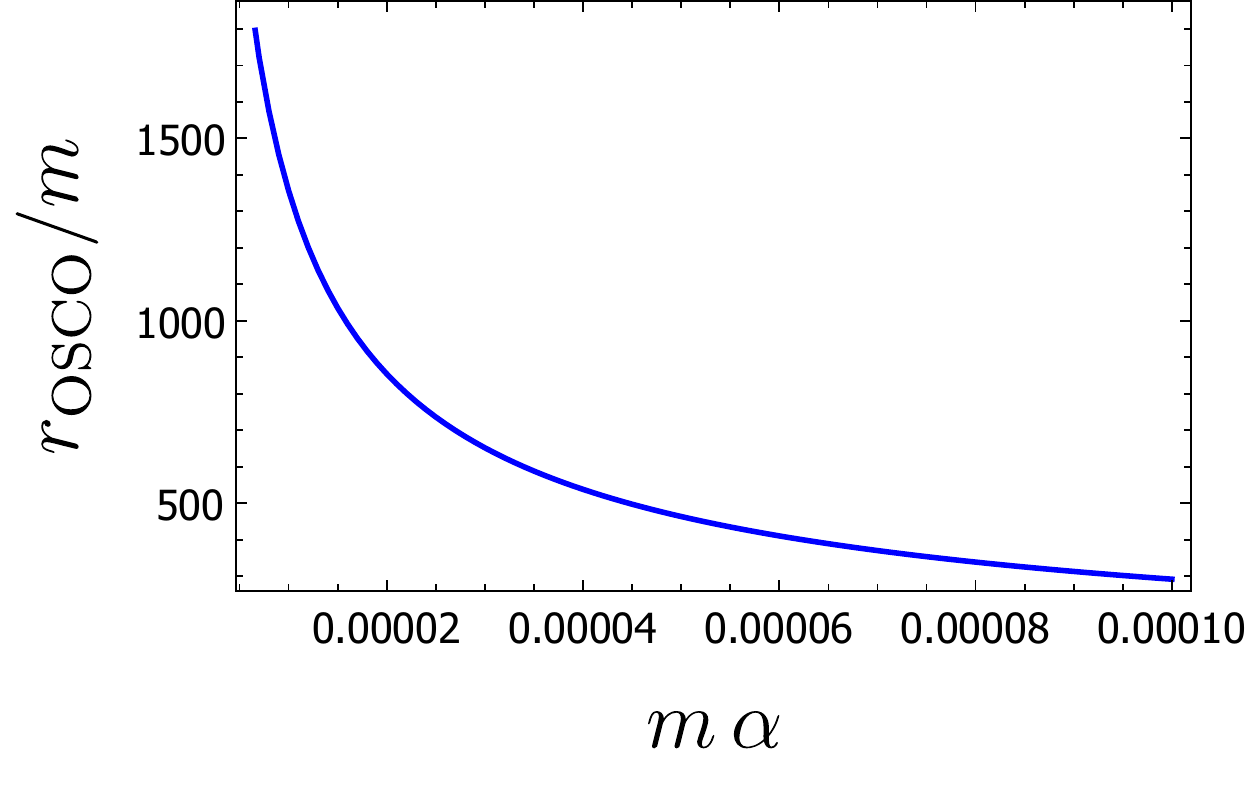}
	\includegraphics[width=0.45\textwidth]{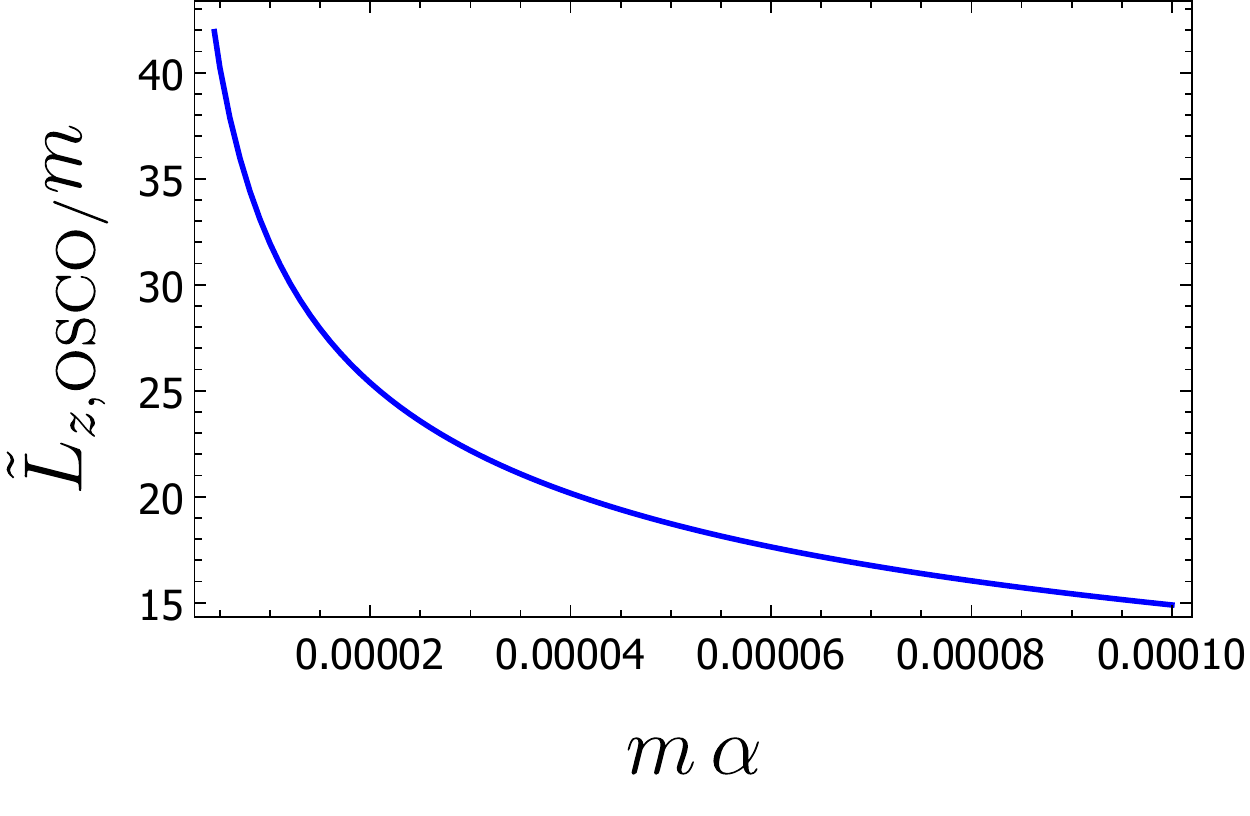}
	\caption{\textit{Top}: The radius of OSCO as a function of the acceleration of the black hole. \textit{Bottom}: The angular momentum of the particle on OSCO. We have taken $m^2 \Lambda = 10^{-52}$.}
	\label{fig:rlosco}
\end{figure}

In Fig.~\ref{fig:rlosco} we see that the radius of OSCO and the angular momentum of particle at this orbit decrease by increasing the acceleration. We also note that there is a critical value of acceleration for which OSCO and ISCO would be on top of each other. (For the value of the cosmological constant taken in Figs.~\ref{fig:isco} and \ref{fig:rlosco} the critical acceleration is $m \alpha \simeq 1.61466 \times 10^{-2}$.)

\begin{figure}[htp]
	\centering
	\includegraphics[width=0.45\textwidth]{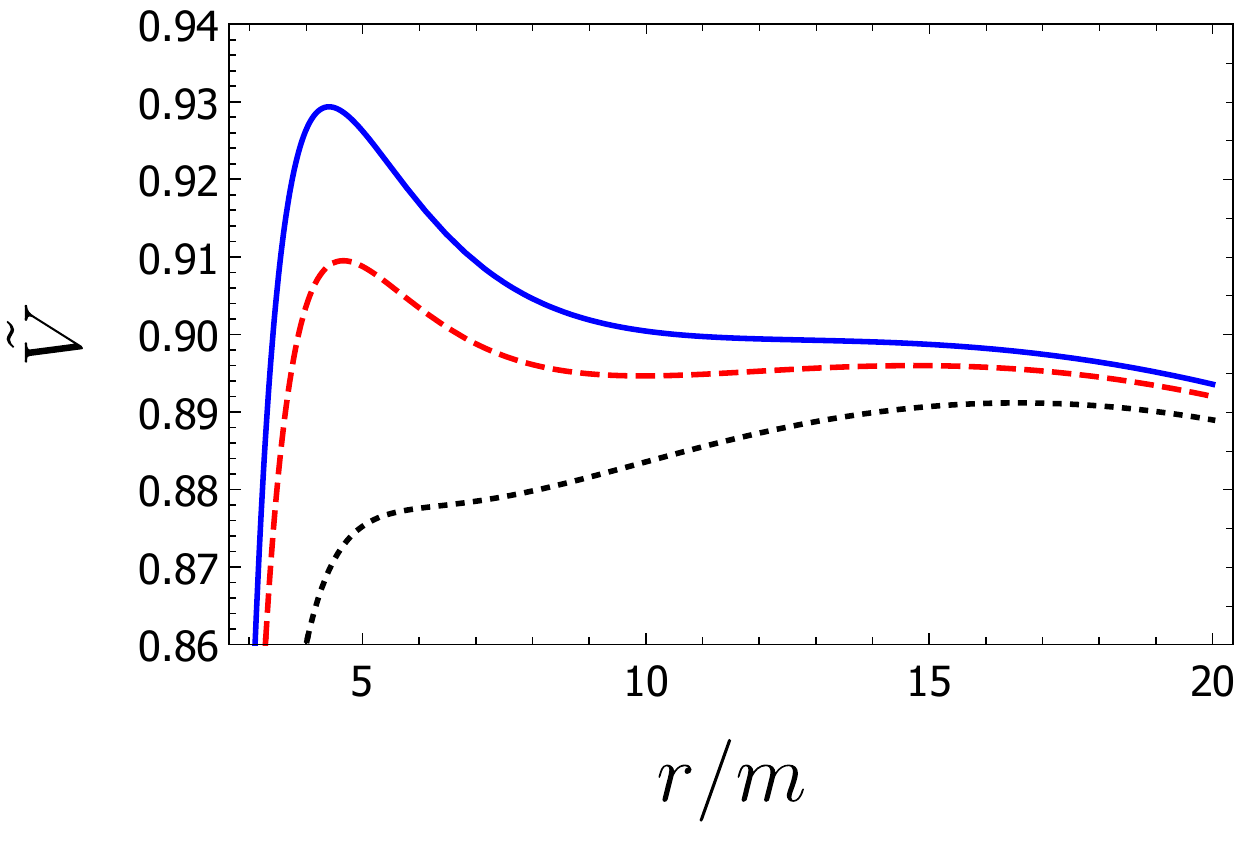}
	\caption{The effective potential for different values of angular momentum in the near and far region from the black hole horizon. We have taken $m \alpha = 10^{-2}$, $\Lambda = 0$, and $\tilde{L}_{z}=3.4 m < \tilde{L}_{z, {\rm ISCO}}$ (dotted black plot), $\tilde{L}_{z, {\rm ISCO}} < \tilde{L}_{z}=3.6 m < \tilde{L}_{z, {\rm OSCO}}$ (dashed red plot), and $\tilde{L}_{z}= 3.7 m > \tilde{L}_{z, {\rm OSCO}}$ (solid blue plot). Qualitatively similar behaviors take place in the de Sitter and anti de Sitter background as well.}
	\label{fig:nearnfar}
\end{figure}

Behavior of the potential in both the near and the far region is presented in Fig.~\ref{fig:nearnfar}. For the case in which $\tilde{L}_{z}<\tilde{L}_{z, {\rm ISCO}}$ ($\tilde{L}_{z}>\tilde{L}_{z, {\rm OSCO}}$) the potential has a maximum in the far (near) region. If the energy of the particle coming from infinity is larger than the maximum value of the potential it will plunge into the black hole. The case in which $\tilde{L}_{z, {\rm ISCO}} < \tilde{L}_{z}< \tilde{L}_{z, {\rm OSCO}}$ is more interesting and has been presented in the middle (dashed red) plot of Fig.~\ref{fig:nearnfar}. In this case there are two maxima at $r_{{\rm max}_1}$ and $r_{{\rm max}_2}$ (assuming $r_{{\rm max}_1}<r_{{\rm max}_2}$ and $\tilde{V}(r_{{\rm max}_1})>\tilde{V}(r_{{\rm max}_2})$) and one minimum between them at $r_{{\rm min}}$.

Ignoring the particles confining to $r<r_{{\rm max}_1}$, depending on the energy of the particle, the following cases could happen:
\begin{itemize}
	\item For a particle approaching the black hole, if the energy satisfies $\tilde{E}^2>\tilde{V}(r_{{\rm max}_1})$, the particle would plunge into the black hole.
	\item If $\tilde{V}(r_{{\rm max}_2})<\tilde{E}^2<\tilde{V}(r_{{\rm max}_1})$, the particle reaches a minimum radius in the range $r_{{\rm max}_1}<r<r_{{\rm min}}$ and then escape to infinity.
	\item If $\tilde{V}(r_{{\rm min}})<\tilde{E}^2<\tilde{V}(r_{{\rm max}_2})$, and if the particle is coming from infinity it reaches a minimum radius of $r>r_{{\rm max}_2}$. The other possibility is that the particle rotates around the black hole in a bound orbit with radius $r_{{\rm max}_1}<r<r_{{\rm max}_2}$.
	\item If $\tilde{E}^2<\tilde{V}(r_{{\rm min}})$ the particles coming from infinity would reach a minimum at $r>r_{{\rm max}_2}$ (here $r$ is larger than the case in which $\tilde{V}(r_{{\rm min}})<\tilde{E}^2<\tilde{V}(r_{{\rm max}_2})$) and then escape to infinity.
\end{itemize}
Also if $\tilde{E}^2=\tilde{V}(r_{{\rm max}_1})$ or $\tilde{E}^2=\tilde{V}(r_{{\rm max}_2})$ the particle will rotate the black hole on an unstable circular orbit, and if $\tilde{E}^2=\tilde{V}(r_{{\rm min}})$ the particle rotate the black hole on a stable circular orbit.

\begin{figure}[htp]
	\centering
	\includegraphics[width=0.45\textwidth]{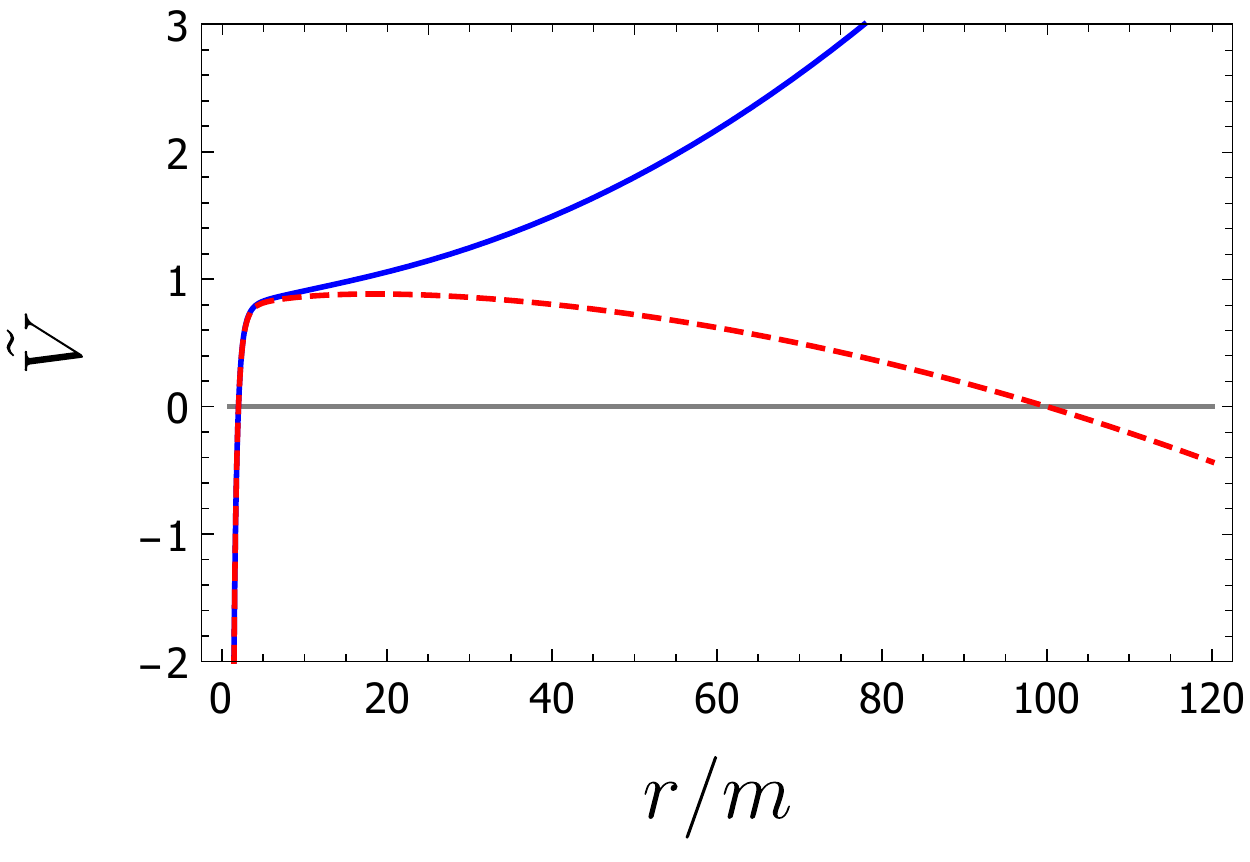}
	\caption{The effective potential for $m \alpha = 10^{-6}$ and $m^2 \Lambda = -10^{-3}$ (solid blue plot) and $m \alpha = 10^{-2}$ and $m^2 \Lambda = -10^{-10}$ (dashed red plot). The solid blue plot satisfies $\Lambda < -3 \alpha^2$, while for the dashed red plot $\Lambda > -3 \alpha^2$. We have taken $\tilde{L}_{z}=3 m$ in both plots.}
	\label{fig:pot_ds_ads}
\end{figure}

We also note that for $\Lambda < -3 \alpha^2$ the potential goes to $+\infty$ as $r$ increases. However for $\Lambda > -3 \alpha^2$ (whether in de Sitter, flat, or anti de Sitter background) the potential goes to $-\infty$ as $r$ increases. These point have been illustrated in Fig.~\ref{fig:pot_ds_ads}.

\section{Precession of perihelion}\label{sec:precession}

In this section we investigate the precession of perihelion for orbits around accelerating black holes. Eq.~\eqref{eqn:lag} can be written as
\be\label{eqn:drdphi}
\left(\frac{dr}{d\phi}\right)^2=\frac{r^4\tilde{E}^2}{\tilde{L}_z^2}-\frac{Qr^4}{\tilde{L}_z^2}-Qr^2.
\ee
At the perihelion and aphelion we have $dr/d\phi=0$. If we denote the radius of the perihelion and aphelion respectively by $r_p$ and $r_a$, we can write the energy and angular momentum per unit rest mass as
\ba
\tilde{E}^2&=&\frac{Q(r_a)Q(r_p)\left(r_p^2-r_a^2\right)}{Q(r_a)r_p^2-Q(r_p)r_a^2},\label{eqn:e}\\
\tilde{L}_z^2&=&\frac{r_a^2r_p^2\left[Q(r_p)-Q(r_a)\right]}{Q(r_a)r_p^2-Q(r_p)r_a^2}. \label{eqn:lz}
\ea

As we mentioned earlier the periodicity of $\phi$ is $2C_0\pi$, with $C_0=(1+2\alpha m)^{-1}$.Therefore, the precession can be found, by using Eq.~\eqref{eqn:drdphi}, as
\ba
\Delta\phi&=&2\left[\phi(r_a)-\phi(r_p)\right]-2(1+2\alpha m)^{-1}\pi \nn\\
&=&2\int_{r_p}^{r_a}\frac{dr}{\sqrt{\frac{r^4\tilde{E}^2}{\tilde{L}_z^2}-\frac{Qr^4}{\tilde{L}_z^2}-Qr^2}}-2(1+2\alpha m)^{-1}\pi ,\nn\\ \label{eqn:precession}
\ea
where $\tilde{E}^2$ and $\tilde{L}_z^2$ are given by eqs.~\eqref{eqn:e} and \eqref{eqn:lz}.

The radius of the perihelion and aphelion are related to the eccentricity $e$ and semi-major axis $a$ of the orbit through $r_p=(1-e)a$ and $r_a=(1+e)a$. As an example we take the semi-major axis and eccentricity of S2 in its orbit around Sgr A*. They are $a=1.543\times10^{14} {\rm m}$ and $e=0.88$ \cite{eisenhauer2003geometric}.

\begin{figure}[htp]
	\centering
	\includegraphics[width=0.45\textwidth]{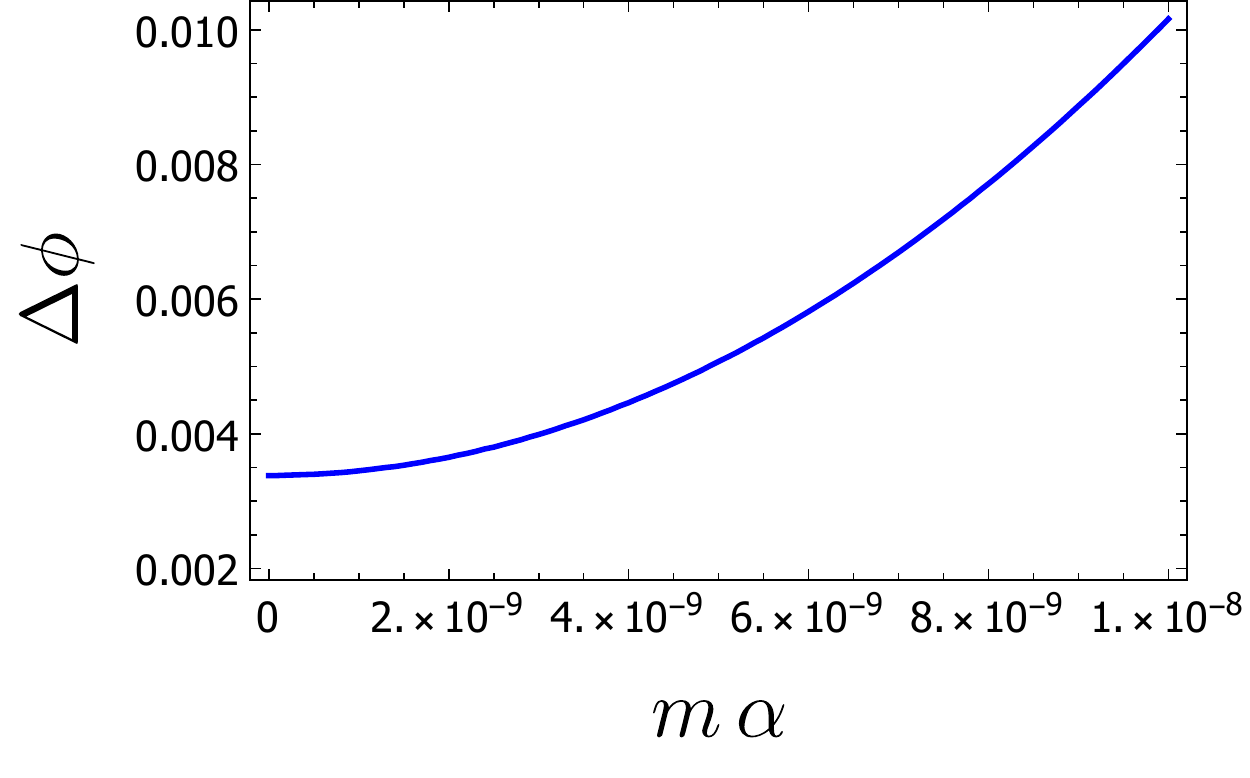}
	\includegraphics[width=0.45\textwidth]{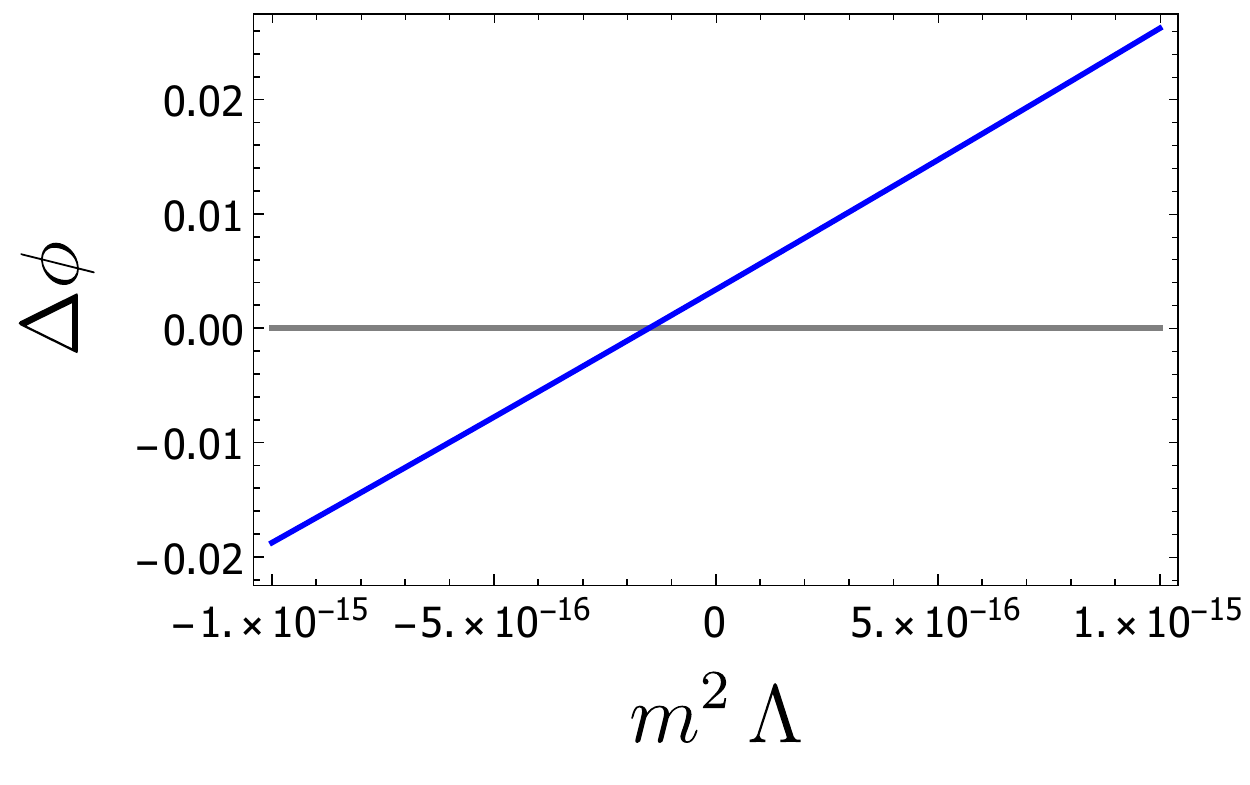}
	\caption{\textit{Top}: Precession of perihelion as a function of the acceleration. We have taken $\Lambda = 0$. \textit{Bottom}: Precession of perihelion as a function of the cosmological constant. We have taken $\alpha = 0$.}
	\label{fig:deltaphi}
\end{figure}

In the top panel of Fig.~\ref{fig:deltaphi} we have plotted the precession \eqref{eqn:precession} as a function of acceleration for S2. We see that the precession angle increases by increasing the acceleration of the central black hole. Also in the bottom panel of Fig.~\ref{fig:deltaphi} we see that the precession increase by increasing the cosmological constant. It is very interesting that the precession angle is negative for $m^2\Lambda \lesssim -1.51363\times 10^{-16}$.

\section{Concluding remarks}\label{sec:con}

We have studied the geodesics of massive particles on equatorial plane of accelerating black holes. We have only considered the component of the acceleration which is perpendicular to the plane of the particle motion. If the component parallel to this plane is considerably large, then the angular momentum of the particle is no longer a constant of motion and we would only have one constant of motion (the energy). In that case the geodesic equations cannot be solved analytically.

We have found several new and interesting results. The study of radial geodesics and stable circular orbits around an accelerating black hole show that there exists some sort of similarity between accelerating black hole and black holes in de Sitter background. Radially outward motion of test particles in accelerating black hole spacetime points out to the fact that acceleration, like a positive cosmological constant, generate a repulsive force. (However, in the case of de Sitter space, the ``cosmic acceleration'' is spherically symmetric, whereas here in acceleration is in $z$ direction.) Another similarity is the existence of OSCO in accelerating black hole spacetime which also exists for particles orbiting a black hole in de Sitter background.

We have found that the precession of perihelion is larger around an accelerating black hole. It is also shown that in anti de Sitter background the precession can be negative. We postpone a detailed study of the precession in anti de Sitter background to a future work. It would also be very interesting to study the timelike geodesics around rotating accelerating black holes.

\section{Acknowledgments}

We are grateful to M.M. Sheikh-Jabbari for reading several pre-publication versions of this paper and providing us with useful comments.
We also acknowledge the support of Iran Science Elites Federation and the hospitality of the University of Guilan.
We would also like to thank B. Turimov and O. Yunusov for pointing out the mistake in the plots on the top panels of Fig.~\ref{fig:isco} and Fig.~\ref{fig:rlosco} in the earlier versions of this paper.

\bibliography{mybib}
\end{document}